\documentclass[prl,twocolumn]{revtex4}
\usepackage{graphicx}
\usepackage{epsfig}
\usepackage{dcolumn}
\usepackage{bm}
\usepackage[latin1]{inputenc}
\begin{document}
\title{Modeling Equilibrium Clusters in Lysozyme Solutions}
\author{Fr\'{e}d\'{e}ric Cardinaux\footnote[1]{current address: Physik der weichen
Materie, IPkM, Heinrich-Heine-Universit\"{a}t D\"{u}sseldorf, 40225
D\"{u}sseldorf, Germany.}, Anna Stradner, Peter Schurtenberger}
   \affiliation{Department of Physics, University of Fribourg, CH-1700 Fribourg, Switzerland.}
\author{Francesco Sciortino, Emanuela Zaccarelli}
    \affiliation{Dipartimento di Fisica and CNR-INFM-SOFT, Universit\`{a} di Roma La Sapienza, Italy.}
\date{\today}
\begin{abstract}
We present a combined experimental and numerical study of the
equilibrium cluster formation in globular protein solutions under
no-added salt conditions. We show that a cluster phase emerges as a
result of a competition between a long-range screened Coulomb
repulsion and a short-range attraction. A simple effective
potential, in which only depth and width of the
attractive part of the potential
are optimized, accounts in a remarkable way for the wavevector
dependence of the X-ray scattering structure factor.
\end{abstract}
\maketitle

Competition between short-range attraction and long range repulsion
provides an efficient way to stabilize aggregates whose shape and
characteristic size result from the delicate balance between these
opposing
forces\cite{1999JChPh.Sear,GroenewoldJ._jp011646w,2004PhRvL.Sciortino,2004Natur.Stradner,2005PhRvL.Campbell,2005PhRvE.Bordi}.
Under appropriate external conditions, particles interacting with
such a mixed potential may form equilibrium cluster phases. This is
a state of matter in which the stable structure of the solution is
characterized by the presence of equilibrium aggregates of
particles, the colloid analog of micelles\cite{Wud92a}.

Cluster phases have been recently observed in colloidal systems as a
result of the competition between short-range depletion attraction
and long-ranged electrostatic
repulsion\cite{2005PhRvL.Campbell,2004Natur.Stradner}. Confocal
microscopy has provided detailed information on the size and shape
of these clusters. Theoretical and numerical studies suggest that
cluster-phases can also be observed in different systems such as
star-polymer solutions\cite{Stiak05} or charged
lyposomes\cite{2005PhRvE.Bordi}. The typical signature of the
equilibrium cluster phase is a pre-peak in the structure factor,
signaling a preferential distance set by the competing forces on
different length scales. Such a feature has been recently reported
in solutions of globular proteins\cite{2004Natur.Stradner,Bagl04},
implying a generality of the mechanism by which bulk aggregation is
disfavored
and finite-size clusters are formed and persist in
equilibrium. This similarity suggests the possibility  of an
approach to
globular proteins based on the assumption of an
effective interaction potential, in which the competition between
repulsion and attraction is built in.

In proteins, short-range attraction is normally attributed to a
combination of van der Waals attraction, hydration forces, and
hydrophobic interactions\cite{1997.Ducruix,Sear06}. While in
colloid-polymer mixtures such short-range attraction is well
characterized in terms of polymer size and concentration, in protein
solutions it is poorly understood. However, clear evidence for its
presence is provided by studies at increasing ionic strength of the
second virial coefficient
\cite{1996Rosenbaum,1998PhRvE.Piazza,2003PRE.Allahyarov} and by the
determination of gas-liquid coexistence lines
\cite{1987Thomson,1996PhRvEBroide,Mus97},
which are metastable with respect to crystallization as for
short-ranged attractive colloids\cite{And02a}.  Long-range repulsion
arises from screened electrostatic interactions, associated to the
net charge of the protein in pH-controlled protein solutions.

Among globular proteins, lysozyme has become the prototype for
scientific investigations.  Under no-added salt condition, the
lysozyme structure factor shows a clear cluster pre-peak whose
position is essentially independent of protein concentration and
weakly dependent on temperature\cite{2004Natur.Stradner}. Hence, it
is particularly important to assess under which conditions an
effective potential can be designed which accounts for such typical
features. Previously, theoretical studies have attempted to study
lysozyme solutions for different solution parameters (e.g. pH,
density, salt concentration), either with an effective continuum
model for electrostatic interactions between proteins carrying
discrete charges\cite{2001JPCB.Carlsson} or with explicit primitive
models\cite{2003PRE.Allahyarov}. At high ionic strength, modeling
for the calculation of the phase diagram usually relies on purely
attractive potentials\cite{Pell03}. Some models have incorporated
the `patchyness' and/or the non-sphericity of the
interactions\cite{2001JPCB.Carlsson,bened99}, associated with the
discrete distribution of charged and hydrophobic sites, whose
relative distances could compete with the relevant distances
incorporated in an effective spherical potential.  But, to our
knowledge, for none of these models, despite the large number of
involved parameters, an equilibrium cluster phase has been
predicted.

In this Letter we present a combined experimental and numerical study
on the equilibrium cluster formation in suspensions of a globular
protein in the absence of added salt.
We propose an effective simple
potential for the lysozyme-lysozyme interaction in water
based on a short-range attractive potential complemented by a Yukawa
screened electrostatic repulsion. Parameters in the Yukawa potential
are fixed by the known size and charge of the protein and the
composition of the solvent. In contrast to previous studies on
colloid-polymer mixtures where the contribution of background ions
is dominant\cite{2003JPCM.Royall,2005JPCB.Sciortino}, counterions
coming from the surface charge induce a highly
concentration-dependent behavior of the amplitude and screening
length of the Yukawa potential, which is properly taken into
account\cite{1986JChPh.Belloni}. The depth and width of the
potential is chosen by best-fit with the structure factor
at one reference concentration, and kept fixed for all other studied
state points. The resulting potential is capable of accurately
describing the measured static properties at low and intermediate
concentrations.

The globular protein used is hen egg white lysozyme (Fluka, L7651).
Its molecular weight is $14.4$kDa and its shape ellipsoidal with
linear dimensions of $3\times3\times4.5$nm\cite{1996PhRvEBroide}.
For simplicity, we neglect the
asymmetry in shape, and we model it as a sphere of diameter
$\sigma=3.4$nm. Lysozyme is dispersed in a solution of D$_{2}$O
(99.9\%, Cambridge Isotope Laboratories) containing 20mM Hepes
buffer salt. A detailed description of the sample preparation
procedure can be found elsewhere\cite{2004Natur.Stradner}. The
$pH=7.8$ is adjusted with sodium hydroxide and held constant within
$\pm0.1$ units for all volume fractions investigated. Under these
conditions, the net charge of the protein is known from titration
experiments to be $Z_0=+8e$ \cite{1972Biochem.Tanford}. The ionic
strength of the solvent, estimated by conductimetry, is $8$mM,
corresponding to a Debye screening length at infinite protein
dilution of $\xi\simeq3.4$nm. SAXS experiments are performed with a
pinhole camera (NanoSTAR from Bruker AXS) equipped with a sealed
tube (Cu K$_{\alpha}$), a thermostatically regulated sample chamber
and a two-dimensional gas detector. Samples are measured at volume
fractions ranging from $\phi=0.085$ to $0.201$, where
$\phi=\pi\rho\sigma^3/6$, with $\rho$ the protein number density, is
systematically measured by UV-visible spectroscopy.

\begin{figure}[t]
\begin{center}
\epsfig{file=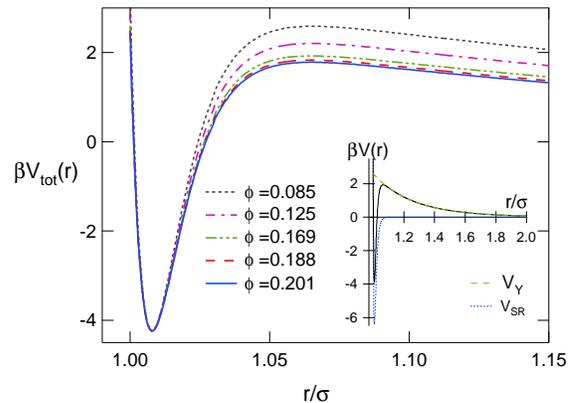, width=220pt}
 \caption{Potential $V_{tot}(r)$  as a function of
 interparticle distance $r$ for various $\phi$. The potential well depth is
 $4.22k_BT$ and its range is $\approx3.6\%\sigma$. The inset details the construction of $V_{tot}$
 with a short-range attractive potential $V_{SR}$ and a long-range
 screened Coulomb repulsion $V_Y$.} \label{Graph1}.
 \end{center}
\vspace{-1.1cm}
\end{figure}

We perform molecular dynamics simulations (MD) of a system composed
of $N=2500$ particles of diameter $\sigma$ and mass $m$ in a cubic
box of size $L$, as a function of volume fraction $\phi$
and  temperature $T$. The excluded-volume term plus the short-range
attraction are modeled for simplicity with the generalized
Lennard-Jones $2\alpha-\alpha$ potential
\cite{1999PhyA.Vliegenthart},
\begin{equation} V_{SR}(r)=4 \epsilon \bigg[
\bigg(\frac{\sigma}{r}\bigg)^{2\alpha}- \bigg(\frac
{\sigma}{r}\bigg)^{\alpha}\bigg], \label{eq:potsr}
\end{equation}
where $\alpha$ essentially controls the width of the attraction and
the steepness of the hard-core, while $-\epsilon$ is the depth of
the potential. The parameters $\sigma$ and $\epsilon$ are chosen as
units of length and energy respectively. The short-range potential
is complemented by a long-range repulsion, modeled with a Yukawa
potential.  Its amplitude and screening length are fixed by the
experimental conditions and follow the generalized one-component
macroion model (GOCM)\cite{1986JChPh.Belloni,Belloni2000},
\begin{equation}
   V_Y(r)=k_B T L_B Z^2_0  X^2 \frac{e^{-r/\xi}}{r}, \label{eq:gocm}
\end{equation}
where $L_B=e^2/(4\pi \varepsilon_0 \varepsilon k_B T)$ is the
Bjerrum length, $\xi=4\pi L_B (\rho Z_0+2\rho_s)^{-1/2}$ is the
Debye length, with $\rho$ the protein number density, $Z_0$ the net
charge on a protein, $\rho_s$ the salt number density in the buffer
and $X$ a correction factor that depends on both $\phi$ and $\xi$
(Eq.  (11)-(15) in \cite{1986JChPh.Belloni}). The resulting Yukawa
potential gives a realistic description of the effective repulsion
between proteins for the relatively high volume fractions
investigated. Here the explicit $\phi$-dependence of $X$
incorporates the effect of screening of a protein by other proteins,
whereas the contribution of additional counter-ions with increasing
$\rho$ enters in the calculation of $\xi$. Under the chosen
condition of low background electrolyte, we find it important to
properly describe the change of $\xi$ and $X$ with $\phi$. Under
excess salt, or infinite dilution, the GOCM reduces to the repulsive
part of the classical DLVO potential. Particles are assumed to
interact simultaneously via $V_{tot}=V_{SR}+V_Y$, shown in
Fig.~\ref{Graph1} for various $\phi$ values.
The integration time-step is fixed to $\Delta t=5~10^{-3}$ in units
of $\sqrt{m \sigma^2 / \epsilon}$.  A cutoff at $r_c=8 \xi$ is
applied to reduce the computational effort, without significantly
altering the model, since $V_{tot}(r_c) < 10^{-3}$.

Whilst the theoretical work of Belloni allows us to model the
electrostatic repulsion without any fitting parameter, no theory is
capable to properly account for the attractive part.
However, in the case of lysozyme and numerous other globular
proteins, attraction is known to be short-ranged and of moderate
strength\cite{Pia00a}. Previous scattering studies on lysozyme
suspensions under high salt conditions have shown that the
attractive part of the effective potential does not significantly
depend on
temperature\cite{1998PhRvE.Piazza,1996JCG.Rosenbaum,1996JChPh.Malfois}.
Following these indications, the
depth and width
of the Lennard-Jones potential in Eq.  (\ref{eq:potsr}) are chosen
by running simulations at fixed $\phi=0.188$ for various $\alpha$,
and searching for the $\alpha$-$\epsilon$ values best matching the
experimental peak positions of $S(q)$ at $5^\circ$C.

\begin{figure}[t]
\begin{center}
\epsfig{file=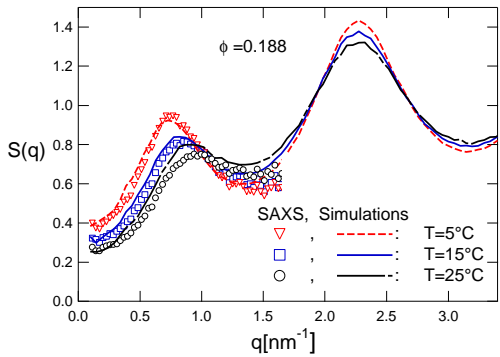, width=200pt}
 \caption{Comparison of the structure factors $S(q)$ obtained by MD
 simulations (lines) with SAXS (symbols), for lysozyme
 suspensions of volume fraction $0.188$ at corresponding temperatures
 for $\alpha=90$.}\label{Graph2}.
 \end{center}
\vspace{-1cm}
\end{figure}
The results from simulations obtained with $\alpha=90$,
corresponding to an attraction range of $3.6\%\sigma$ and
$\epsilon=160$m$e$V or $\approx 6.5k_BT$, are compared with the
measured $S(q)$ in Fig. \ref{Graph2} at $\phi=0.188$.  The position
of the cluster peak and its $T$-dependence are properly accounted
for the chosen $\alpha$ and $\epsilon$ values. It is important to
point out that also the location of the nearest neighbor peaks at
$\approx 2.25$ nm$^{-1}$ coincides with previous SANS
measurements\cite{2004Natur.Stradner}. We also note that the low
$q$-limit is rather well described, suggesting that the present
potential properly accounts for the system compressibility ($\propto
S(0)^{-1}$) for all temperatures. This remarkable agreement infers
that the use of an effective potential captures the essential
features necessary to describe the clustering process in lysozyme
suspensions.

\begin{figure}[t]
\begin{center}
\epsfig{file=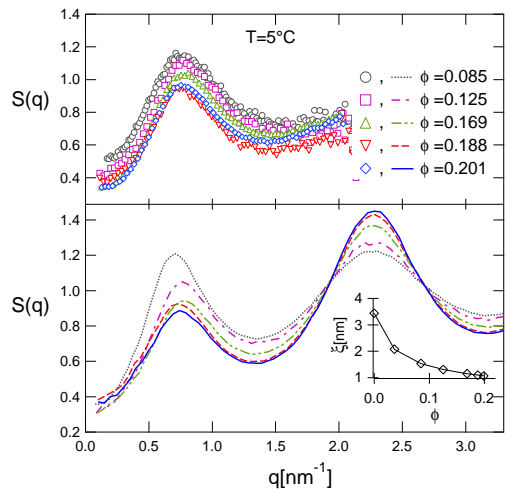, width=200pt} \caption{$\phi$-dependence of
the static structure factor $S(q)$ measured by SAXS (upper panel)
and obtained by MD simulation (lower panel) with potentials of
Fig.\ref{Graph1}. The dependence of the Debye length $\xi$ on volume
fraction is shown as inset.}\label{Graph3}.
\end{center}
\vspace{-1.0cm}
\end{figure}

To validate the resulting potential,  we turn to examine other
densities. In Fig.~\ref{Graph3} numerical results are compared with
the  corresponding experimental data at $T=5^\circ$C for volume
fractions ranging from $\phi=0.085$ to $0.201$.  It is important to
stress that the same well depth and range in the mixed potential is
used at all investigated $\phi$: the only $\phi$-dependence in the
potential arises from the repulsive part, where both $\xi$ and $X$
vary (see Fig.\ref{Graph1}). Again, the numerical $S(q)$ properly
reproduces the features observed experimentally. With increasing
$\phi$, the cluster peak position does not significantly change
while its amplitude systematically decreases. Moreover, the
agreement is quantitative for the cluster and nearest-neighbor peak
positions as well as for their amplitudes, albeit slightly
overestimated at the lowest $\phi$. Also the compressibilities are
well reproduced, with an initial decrease with increasing $\phi$,
saturating to a roughly constant value.

\begin{figure}[t]
\begin{center}
  \epsfig{file=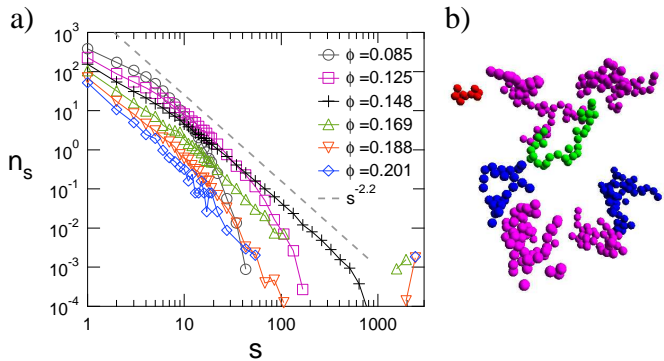, bbllx=40, bblly=30, bburx=703, bbury=350, width=260pt}
 \caption{a) Cluster size distribution $n(s)$ for various $\phi$ at
 $T=5°$C. The distribution for small sizes follows random
 percolation dependence of $s^{-2.2}$. For $\phi>0.148$ large clusters
 are found indicating that the system is percolating. b) Snapshot at $\phi=0.125$ with few  selected clusters of size $n=10, 30, 82, 203$.}\label{Graph4}.
 \end{center}
\vspace{-1cm}
\end{figure}

The ability of the potential to correctly reproduce the structure of
the system suggests to analyze the equilibrium configurations
provided by MD. Particles are considered to be in the same cluster
when the separation between pairs of nearest-neighbor particles is
smaller than $r_{max}=1.065\sigma$, a value
corresponding to the distance at the maximum in $V_{tot}$ (see
Fig.\ref{Graph1}). Fig.~\ref{Graph4}a) shows the resulting
distribution $n(s)$ of clusters of size $s$, for several state
points. We note that these results are qualitatively unaffected by
different choices of the bonding distance. At the smallest $\phi$,
the suspension is essentially composed of small clusters with a
distribution that rapidly and monotonically decays to zero. As
$\phi$ increases, larger clusters form and the distribution
progressively develops a power law behavior $n(s)\propto s^{\tau}$,
with an exponent $\tau\approx -2.2$ consistent with random
percolation.

We can try to exploit the micelle analogy in order to understand the
strong variation of the cluster size distribution shown in
Fig.\ref{Graph4}. In the case of charged micelles at low ionic
strength, one expects a dramatic change in the micellar growth
scenario at a crossover concentration where the screening length
becomes comparable to the micellar length \cite{1990MacKin}. At this
point, a sharp transition to accelerated micellar growth with a very
broad size distribution occurs. In our case, the crossover is
reached once $\xi/\sigma\approx 0.5$. For smaller $\xi$, we then
observe enhanced cluster growth and a broadening of the size
distribution.

For $\phi>0.148$ percolating states are found where the system forms
a space-spanning cluster. From the analysis of the MD trajectories
one observes that clusters are highly transient.
Even above the percolation threshold, the system is still in a fluid
phase, and percolation does not imply gelation, for which a
long-living
network is necessary. Interestingly, the observed transient
percolating states resemble those of the transient networks found in
the semidilute regime of charged micelles, where under the same
salt-free conditions a persisting low-q peak is also
present\cite{1998Lang.Oberdisse}. The observation of isolated
clusters, as depicted in Fig.\ref{Graph4}b), reveals that the
clusters are highly random. We find both space-filling spherical
shapes as well as filamentous structures, made primarily of single
chains, that are very different from those observed in other more
repulsive systems\cite{2005PhRvL.Campbell,2005JPCB.Sciortino}.
Moreover, in the investigated region of phase space, there is not a
preferred cluster size, since a peak in $n(s)$ is never observed.


Clusters observed under the present no-salt conditions are generated
by the competition between attraction and repulsion, as clearly
indicated by the presence of the pre-peak in the structure factor
and should not be considered as the transient clusters that are
commonly observed on approach to a phase-separation boundary in
short-range potentials, whose spectral signature is a peak in $S(q)$
at $q \rightarrow 0$.
In the investigated $T$-range, attraction is always sufficiently
counter-balanced by the electrostatic repulsion, allowing for the
existence of stable clusters and preventing phase separation.
The resulting clusters are highly polydisperse and their structural
development with volume fraction leads to transient networking, with
hallmarks of random percolation.

In summary, we demonstrated that equilibrium cluster formation in
complex protein solutions can be reproduced using a unifying
"colloid-approach", with a simple one-component effective potential
between proteins composed of a short-range attraction and a longer
range repulsion.
This indicates the existence of a fundamental principle for
self-assembly in biological solutions.


We acknowledge support from MIUR-Firb, MIUR-Cofin, the Swiss National Science
Foundation, the State Secretariat for Education and Research (SER)
of Switzerland, and the Marie Curie Network on Dynamical Arrest of
Soft Matter and Colloids (MCRTN-CT-2003504712).

\bibliography{ecf}

\end{document}